\documentclass[preprint,12pt]{elsarticle}
\usepackage{graphicx}
\usepackage{amsmath,amssymb}

\DeclareMathOperator{\tr}{tr} 

\begin{document}

\begin{frontmatter}

\title{Turing instabilities in reaction-diffusion systems with cross diffusion}
\author[energetica, infn]{D. Fanelli}
\author[sistemi, infn]{C. Cianci}
\author[energetica, infn]{F. Di Patti\corref{cor1}\fnref{fn1}}
\ead{f.dipatti@gmail.com}
\address[energetica]{Dipartimento di Energetica, Universit\`{a} degli Studi di Firenze, via S. Marta 3, 50139 Florence, Italy}
\address[infn]{INFN, Sezione di Firenze, via G. Sansone 1, 50019 Sesto Fiorentino, Florence, Italy}
\address[sistemi]{Dipartimento di Sistemi e Informatica, Universit\`{a} degli Studi di Firenze,  via S. Marta 3, 50139 Florence, Italy}
\cortext[cor1]{Corresponding author}
\fntext[fn1]{Tel. +390554796208}
\begin{abstract}
The Turing instability paradigm is revisited in the context of a multispecies diffusion scheme derived from a self-consistent microscopic formulation. The analysis is developed with reference to the case of two species. These latter share the same spatial reservoir and experience a degree of mutual interference due to the competition for the available resources. Turing instability can set in for all ratios of the main diffusivities, also when the (isolated) activator diffuses faster then the (isolated) inhibitor. This conclusion, at odd with the conventional vision, is here exemplified for the Brusselator model and ultimately stems from having assumed a generalized model of multispecies diffusion, fully anchored to first principles, which also holds under crowded conditions.
\end{abstract}

\end{frontmatter}

\section{Introduction}
Turing instability is one of the reference mechanisms for pattern formation in nature \cite{turing1952,buceta2002}. The Turing idea applies to a large gallery of phenomena that can be modelled via reaction-diffusion equations \cite{murray,maynard1974}. These latter are mathematical models that describe the dynamical evolution of distinct families of constituents, mutually coupled and freely diffusing in the embedding medium. Diffusion can seed the instability by perturbing the mean--field homogeneous state, through an activator--inhibitor mechanism, and so yielding the emergence of patched, non homogeneous in space, density distributions. The most intriguing applications of the Turing paradigm are encountered in the context of morphogenesis, the branch of embryology which studies the development of patterns and forms in biology. The realm of application of the Turing ideas encompasses however different fields, ranging from chemistry \cite{kepper,leng,vanag} to biology \cite{murray,levin1976,mimura1978,baurmann2007,wilson1999,shiferaw2006,kikla}, passing through physics \cite{astrov2001}, where large communities of homologous 
elements evolve and interact.

According to the classical viewpoint, however, the diffusion coefficient of the inhibitor species has to be larger than that of the activator, for the patterns to eventually develop. This is a strict mathematical constraint which is not always met in e.g. contexts of biological relevance \cite{rev,strier}, and which limits the possibility of establishing a quantitative match between theory and empirical data. Spatially extended systems made of interacting species sharing similar diffusivities can indeed display self-organized patched patterns, an observation that still calls for a sound interpretative scenario, beyond the classical Turing mechanisms \cite{turing1952}. 

One viable strategy to possibly reconcile theory and observations has been explored in \cite{butler2009} and \cite{biancalani2010}. In these studies, the authors considered the spontaneous emergence of persistent spatial patterns as mediated by the demographic endogenous  noise, stemming from the intimate discreteness of the scrutinized system. The intrinsic noise translates into a systematic enlargement of the parameter region yielding  the Turing order, when compared to the corresponding domain predicted within the deterministic linear stability analysis. It is however unclear at present whether experimentally recorded patterns bear the imprint of the stochasticity, a possibility that deserves to be further challenged in the future.

Alternatively, and to bridge the gap with the experiments, the Turing instability concept has been applied to generalized reaction--diffusion equations. These latter account for cross diffusion terms which are hypothesized to exist on purely heuristic grounds or by invoking the phenomenological theory of linear non--equilibrium thermodynamics \cite{deG84,cross1,cross2}. Diagonal and off--diagonal coefficients of the diffusion matrix are not linked to any microscopic representation of the examined dynamics and are hence treated as free parameters of the model. In \cite{kumar2011} the authors quantify the impact of cross terms on the Turing bifurcation,  showing e.g that spatial order can  materialize also if the inhibitor's diffusion ability is less pronounced than the activator's one. 

Starting from this setting, the aims of this paper are twofold. On the one side, we shall elaborate on a microscopic theory of multispecies diffusion, fully justified from first principles. The theory here derived is specifically targeted to the two species case study and extends beyond the formulation of \cite{fanellimckane}. On the other side, and with reference to the Brusselator model, we will show that Turing patterns can take place for any ratio of the main diffusivities. In doing so we will cast the conclusions of \cite{kumar2011} into a descriptive framework of broad applied and fundamental interest, where the key cross diffusion ingredients are not simply guessed a priori but rigorously obtained via a self--consistent derivation anchored to the microscopic world. Working in the context of a reference case study, the Brusselator model, we shall also perform numerical simulations based on both the underlying stochastic picture and the idealized mean--field formulation to elaborate on the robustness of the observed patterns. 

In the following we briefly discuss the derivation of the model, focusing on the specific case where 
two species are supposed to diffuse, sharing the same spatial reservoir.

\section{The model}
Consider a generic microscopic system bound to occupy a given volume of a $d-$dimensional space. Assume the volume to be partitioned into a large number $\Omega$ of small hypercubic patches, each of linear size $l$. Each mesoscopic cell, labelled by $i$, is characterized by a finite carrying capacity: it can host up to $N$ particles, namely $n^{A}_i$ of type $A$, $n^{B}_i$ of type $B$, and $v_i = N-n^{A}_i - n^{B}_i$ vacancies, hereafter denoted by $V$. 
In general, the species will also interact, as dictated by specific reaction terms. Let us start by solely focusing on the diffusion part, silencing any direct interaction among elementary constituents. As we shall remark, there exists an indirect degree of coupling that results from the competition for the available spatial resources. In practice, 
the mobility of the particles is balked if the neighbouring patches have no vacancies. Particles may jump into 
a nearest--neighbour patch, only if there is a vacancy to be eventually filled. 
This mechanism translates into the following chemical equation
\begin{eqnarray}
A_i + V_j &&\stackrel{\mu^A}{\longrightarrow} V_i + A_j , \label{mig} \\ 
B_i + V_j &&\stackrel{\mu^B}{\longrightarrow} V_i + B_j  \nonumber
\end{eqnarray}
where $i$ and $j$ label nearest--neighbour patches. Here, $A_i$ and $ B_i$ identify the particles that belong to cell $i$. $V_i$ labels instead the empties that are hosted in patch $i$. The parameters $\mu^{A}$ and $\mu^{B}$ stand for the associated reaction rates. Similar reactions control the migration from cell $j$ towards cell $i$.

In addition, and extending beyond the scheme proposed in \cite{fanellimckane}, we imagine the following reactions to hold:
\begin{eqnarray}
A_i + B_j &\stackrel{\alpha}{\longrightarrow}&  A_j  + B_i , \label{mig1}\\ 
A_j + B_i &\stackrel{\alpha}{\longrightarrow}&  A_i  + B_j  \nonumber
\end{eqnarray}
which in practice account for the possibility that elements $A_i$ (resp. $A_j$) and $B_j$ (resp. $B_i$) swap their actual positions. 

The state of the system is then specified by the number of $A$ and $B$ particles in each patch, the number of vacancies following from a straightforward normalization condition. Introduce the vector $\textbf{n}=(\textbf{n}_{1},\ldots,\textbf{n}_{\Omega})$, where $\textbf{n}_i=(n^{A}_i,n^{B}_i)$. The quantity $T(\textbf{n}'|\textbf{n})$ represents the rate of transition from state $\textbf{n}$, to another state $\textbf{n}'$, compatible with the former. The transition rates associated with the migration between nearest--neighbour, see Eqs. (\ref{mig}), take the form
\begin{equation}
T(n^{(a)}_{i}-1,n^{(a)}_{j}+1|n^{(a)}_{i},n^{(a)}_{j}) =
\frac{\mu^{(a)}}{z\Omega}\frac{n^{(a)}_i}{N} \frac{N-n^{A}_{j}-n^{B}_{j}}{N}, \quad  a=A,B, 
\label{TRs}
\end{equation}
where we have made explicit in $T(\cdot|\cdot)$ the components that are affected by the reactions. As discussed in \cite{fanellimckane}, the factor $N-n^{A}_{j}-n^{B}_{j}$, reflects the natural request of a finite capacity, and will eventually yield  a macroscopic modification of the Fick's law of diffusion. Moreover, chemical equations (\ref{mig1}) result in the following transition rates: 
\begin{eqnarray}
T(n^{A}_{i}-1,n^{A}_{j}+1,n^{B}_{i}+1,n^{B}_{j}-1 | n^{A}_{i},n^{A}_{j},n^{B}_{i},n^{B}_{j}) &=&
\frac{\alpha}{z\Omega}\frac{n^{A}_i}{N} \frac{n^{B}_j}{N}, \label{TRs1}\\ 
T(n^{A}_{i}+1,n^{A}_{j}-1,n^{B}_{i}-1,n^{B}_{j}+1 | n^{A}_{i},n^{A}_{j},n^{B}_{i},n^{B}_{j}) &=&
\frac{\alpha}{z\Omega}\frac{n^{A}_j}{N} \frac{n^{B}_i}{N} \qquad .\nonumber
\end{eqnarray}

The process here imagined is Markov, and the probability $P(\textbf{n},t)$ to observe the system 
in state $\textbf{n}$ at time $t$ is ruled by the master equation
\begin{equation}
\frac{dP(\textbf{n},t)}{dt} = \sum_{\textbf{n}'\neq\textbf{n}} 
\left[ T(\textbf{n}|\textbf{n}')P(\textbf{n}',t) 
- T(\textbf{n}'|\textbf{n})P(\textbf{n},t) \right],
\label{master}
\end{equation}
where the allowed transitions depend on the state of the system via the above relations. Starting from this microscopic, hence inherently stochastic picture, one can derive a self--consistent deterministic formulation, which exactly holds in the continuum limit. Mathematically, one needs to obtain the dynamical equations that govern the time evolution of the ensemble averages 
$\langle n^{A}_{i} \rangle$ and $\langle n^{B}_{i} \rangle$. To this end, multiply first the
master  Eq. (\ref{master}) by $n^{a}_i$, with $a=A,B$, and sum over all $\textbf{n}$. After an algebraic 
manipulation which necessitates shifting some of the sums by $\pm 1$, one eventually gets
\begin{eqnarray}
\frac{d\langle n^{(a)}_{i} \rangle}{dt} &=& \sum_{j \in i} 
\Big [ \langle T(n^{(a)}_{i}+1,n^{(a)}_{j}-1|n^{(a)}_{i},n^{(a)}_{j}) \rangle \label{pre_macro} \\
&\phantom{=}&
+ \langle T(n^{A}_{i}+1,n^{A}_{j}-1,n^{B}_{i}-1,n^{B}_{j}+1 | n^{A}_{i},n^{A}_{j},n^{B}_{i},n^{B}_{j})\rangle \nonumber \\
&\phantom{=}& - \langle T(n^{(a)}_{i}-1,n^{(a)}_{j}+1|n^{(a)}_{i},n^{(a)}_{j}) \rangle \nonumber \\
&\phantom{=}&
-\langle T(n^{A}_{i}-1,n^{A}_{j}+1,n^{B}_{i}+1,n^{B}_{j}-1 | n^{A}_{i},n^{A}_{j},n^{B}_{i},n^{B}_{j})\rangle \nonumber
\Big ], 
\end{eqnarray}
where the notation $\sum_{j \in i}$ means that we are summing over all patches $j$ which are 
nearest--neighbours of patch $i$. The averages in Eq. (\ref{pre_macro}) are performed explicitly by recalling the 
expression for the transition rates as given in Eqs. (\ref{TRs}) and (\ref{TRs1}).  Replace then the averages of products by the products of averages, an operation that proves exact in the continuum limit $N \to \infty$. By introducing the continuum concentration $(\phi_{A,B})_{i} = \lim_{N \to \infty} \frac{\langle n^{A,B}_{i} \rangle}{N}$, rescaling time by a factor of $N\Omega$ and taking the size of the patches to zero one finally gets\footnote{Use has been made of the discrete Laplacian operator 
$\Delta f_{i} = (2/z) \sum_{j \in i} (f_{j} - f_{i})$, which then turns into the continuum 
operator $\nabla$ when sending to zero the size of the patch and 
scaling the rates $\mu^{A,B}$ and $\alpha$ appropriately.}
\begin{eqnarray}
\frac{\partial \phi_{A}}{\partial t} &=& D_{11} \nabla^2 \phi_{A} 
+ D_{12} \left[\phi_{A} \nabla^2 \phi_{B} - \phi_{B} \nabla^2 \phi_{A} \right],  \nonumber \\
\frac{\partial \phi_{B}}{\partial t} &=& D_{22} \nabla^2 \phi_{B}+D_{21}
\left[ \phi_{B} \nabla^2 \phi_{A} - \phi_{A} \nabla^2 \phi_{B} \right],
\label{pdes}
\end{eqnarray}
where\footnote{From the above expressions, one derives the consistency conditions $\mu_{A}>\alpha$ and $\mu_{B}>\alpha$.} $D_{11,22} \rightarrow l^2 \mu_{A,B}$ and $D_{12,21} \rightarrow l^2 (\mu_{A,B}-\alpha)$.
The above system of partial differential equations for the concentration $\phi_A$ and $\phi_B$ is a slightly modified version of the one derived in \cite{fanellimckane}, this latter being formally recovered when setting $\alpha$ to zero. In the generalized context here considered, the cross diffusion coefficients $D_{12}$ and $D_{21}$ are different, specifically smaller, than the corresponding mean diffusivities $D_{11}$ and $D_{22}$. We emphasize again that the crossed, nonlinear contributions  $\pm (\phi_{A,B} \nabla^2 \phi_{B,A} - \phi_{B,A} \nabla^2 \phi_{A,B})$ stem directly from the imposed finite carrying capacity and, as such, have a specific, fully justified, microscopic origin. The diffusive fluxes that drive the changes in the concentrations $\phi_A$ and $\phi_B$ can be written as:
\begin{eqnarray}
\mathbf{J}_{\phi_A} &=& - D_{11} \left(1- \frac{D_{12}}{D_{11}}\phi_B \right) \mathbf{\nabla} \phi_A - D_{12} \phi_A \mathbf{\nabla} \phi_B \nonumber \\
\mathbf{J}_{\phi_B} &=& -  D_{21} \phi_B \mathbf{\nabla} \phi_A - D_{22} \left(1- \frac{D_{21}}{D_{22}} \phi_A \right) \mathbf{\nabla} \phi_B 
\label{modified_Ficks}
\end{eqnarray}

It is interesting to notice that relations (\ref{modified_Ficks}) enable us to make contact with the field of  
linear non--equilibrium thermodynamics (LNET), a branch of statistical physics which defines the general framework for the macroscopic description of e.g. transport processes. One of the central features of LNET is the relation between the forces, which cause the state of the system to change, and the fluxes,  which are the result of these changes~\cite{deG84}. Within the formalism of LNET the fluxes $\mathbf{J}_{\phi_A}$ and $\mathbf{J}_{\phi_B}$ that rule the diffusion of the two species $\phi_A$ and $\phi_B$ are linearly related to the forces, the gradients of the respective concentrations. The quantities that establish the formal link between forces and fluxes are 
the celebrated Onsager coefficients, postulated on pure heuristic grounds. Interestingly, Eqs. (\ref{modified_Ficks})
provide a self--consistent derivation for the Onsager coefficients, that enters the generalized Fick's scenario here depicted.  

Define $\mathbf{\Phi} = (\phi_A,  \phi_B)$ and $\mathbf{J} = (J_{\phi_A},  J_{\phi_B})$. Then Eqs. (\ref{pdes}) can be written in the compact form:
\begin{equation}
\label{compactpde}
\frac{\partial \mathbf{\Phi}}{\partial t} = - \mathbf{\nabla} \mathbf{J} = \mathbf{\nabla} \mathbf{D}(\mathbf{\Phi}) \mathbf{\nabla} \mathbf{\Phi}
\end{equation}
where the $2 \times 2$ matrix $\mathbf{D}$ reads:
\begin{equation*}
\mathbf{D}(\mathbf{\Phi})  = 
\left(
\begin{array}{cc}
D_{11} \left(1- \frac{D_{12}}{D_{11}}\phi_B \right) &   D_{12} \phi_A  \\
D_{21} \phi_B & D_{22} \left(1- \frac{D_{21}}{D_{22}} \phi_A \right)
\end{array}
\right).
\end{equation*}

A stringent constraint from thermodynamics is that all eigenvalues of the diffusion matrix $\mathbf{D}$ are real and positive. This in turn corresponds to requiring $\tr(\mathbf{D}) > 0$ and $\det(\mathbf{D})>0$. A straightforward calculation yields: 
\begin{eqnarray*}
\tr(\mathbf{D}) &=& D_{11} (1-\phi_B)+ D_{22} (1-\phi_A) + \Delta D (\phi_A+\phi_B) \\ 
\det(\mathbf{D})  &=& D_{11} D_{22} (1 -\phi_A -\phi_B) + \Delta D (D_{11} \phi_A + D_{22} \phi_B) 
\end{eqnarray*}
where $\Delta D \equiv D_{11}-D_{12}= D_{22}-D_{21}$. By definition $\Delta D >0$. Moreover, 
$\phi_A$ and $\phi_B$ are both positive and smaller than one. Hence, $\tr(\mathbf{D}) > 0$ and $\det(\mathbf{D})>0$, 
a result that points to the consistency of the proposed formulation. 

\section{The region of Turing order}
Having derived a plausible macroscopic description for the two components diffusion process, we can now move on by allowing the involved species to interact and consequently consider in the mathematical model the corresponding reaction terms. As an important remark, we notice that these latter can be also obtained as follows the above, rather general, approach that bridges micro and macro realms. First, one need to resolve the interactions among individual constituents, by translating into chemical equations the microscopic processes implicated. These include cooperation and competition effects, as well as the indirect interferences stemming from the finite carrying capacity that we have imposed in each mesoscopic patch. Then, one can recover the deterministic equations for the global concentrations, by operating in the continuum system size limit. In general, Eq. (\ref{compactpde}) is modified into:
\begin{equation}
\label{compactpde2}
\frac{\partial \mathbf{\Phi}}{\partial t} = \mathbf{F}(\mathbf{\Phi}) + \mathbf{\nabla} \mathbf{D} \mathbf{\nabla} \mathbf{\Phi}
\end{equation}
where  $\mathbf{F} = (f_A({\phi_A},{\phi_B}), f_B({\phi_A},{\phi_B}))$. As we have anticipated, the interest of this generalized formulation, resides in that it allows for Turing like patterns in a region of the parameter space that is instead forbidden when conventional reaction--diffusion systems are considered. The novelty of the proposed formulation has to do with the presence of specific cross diffusion terms, which follow a sound physical request, and add to the classical Laplacians, signature of Fickean diffusion. 

Let $\hat{\phi}_A, \hat{\phi}_B$ be the steady state solution of the homogeneous (aspatial) system, namely       
$f_A(\hat{\phi}_A,\hat{\phi}_B) = f_B(\hat{\phi}_A, \hat{\phi}_B) = 0$. The fixed point is linearly stable if the Jacobian matrix $\mathbf{A}$
\begin{equation*}
\mathbf{A} = 
\left(
\begin{array}{cc}
\frac{\partial f_A}{\partial {\phi}_A} & \frac{\partial f_A}{\partial {\phi}_B} \\
\frac{\partial f_B}{\partial {\phi}_A} & \frac{\partial f_B}{\partial {\phi}_B} 
\end{array}
\right),
\end{equation*}
has positive determinant and negative trace. It is worth stressing that the derivatives in matrix $\mathbf{A}$ are evaluated at the homogeneous fixed point. Back to the complete model, a spatial perturbation superposed to the homogeneous fixed point can get unstable if specific conditions are met. Such conditions, inspired to the seminal work by Turing, are hereafter derived via a linear stability analysis. Define $\mathbf{\eta} = \mathbf{\Phi}- \hat{\mathbf{\Phi}}$ and proceed with a linearization of Eq. (\ref{compactpde2}) to eventually obtain:
\begin{equation*}
\frac{\partial \mathbf{\eta}}{\partial t} = \mathbf{A}(\hat{\mathbf{\Phi}})\mathbf{\eta} + \mathbf{D}(\hat{\mathbf{\Phi}}) \mathbf{\nabla}^2 \mathbf{\eta}
\end{equation*}
Going to Fourier space one gets:
\begin{equation}
\label{lineartur}
\frac{d \tilde{\mathbf{\eta}}}{d t} = \mathbf{A}^*(k)\tilde{\mathbf{\eta}} 
\end{equation}
where $\mathbf{A}^*(k) = \mathbf{A}(\hat{\mathbf{\Phi}})- k^2 \mathbf{D}(\hat{\mathbf{\Phi}})$. By characterizing the eigenvalues of the matrix $\mathbf{A}^*$, one can determine whether a perturbation to the homogeneous solution can yield patterns formation. In particular, if one of the eigenvalues admits a positive real part for some values of $k$, then a spatially modulated instability develops. The growth of the perturbation as seeded by the linear instability will saturate due to the non linearities and eventually results in a characteristic pattern associated to the unstable mode $k$. Steady patterns of the Turing type require in addition that the imaginary part of the eigenvalues associated to the unstable mode are zero. In formulae, the Turing instability sets in if there exists a $k$ such that $\tr(\mathbf{A}^*(k))<0$  and  $\det(\mathbf{A}^*(k))<0$. These latter conditions are to be imposed, jointly with the request of a stable homogeneous fixed point ($\tr(\mathbf{A})<0$, $\det(\mathbf{A})>0$), to identify the parameters' values that drive the instability. Alternatively, one can obtain a set of explicit conditions following the procedure outlined below, and adapted from \cite{biancalani2010}. The eigenfunctions of the Laplacian operator are:  
\begin{equation*}
\label{eigenlap}
\bigl(\nabla^2 + k^2\bigr)\:\mathbf W_k(\textbf{r}) = 0, 
\end{equation*}
and we write the solution to Eq. (\ref{lineartur}) in the form:
\begin{equation}
\label{soluz}
\mathbf x(t, \textbf{r}) = \sum_k e^{\lambda t}\:a_k\:\mathbf W_k(\textbf{r}).
\end{equation}
By substituting the ansatz \eqref{soluz} into Eq. \eqref{lineartur}
yields:
\begin{equation*}
e^{\lambda t}\bigl[ \mathbf A-k^2\:\mathbf{D}-\lambda\mathbf{1}\bigr]\mathbf W_k = 0.
\end{equation*}
The above system admits a solution if the matrix $\mathbf A-k^2\:\mathbf{D}-\lambda\mathbf{1}$ is singular, i.e.:
\begin{equation}
\label{det}
\det(\mathbf A-k^2\:\mathbf{D}-\lambda\mathbf{1}) = 0.
\end{equation}

The solutions $\lambda(k)$ of (\ref{det}) can be interpreted as dispersion relations. If at least one of the two solutions displays a positive real part, the mode is unstable, and the dynamics drives the system towards a non--homogeneous configuration in response to the initial perturbation. Introduce the auxiliary quantity $\Gamma$ defined as:
 \begin{multline}
\Gamma=D_{11} \frac{\partial f_B}{\partial {\phi}_B} + D_{22} \frac{\partial f_A}{\partial {\phi}_A} -
\hat{\phi}_A \left[ D_{21} \frac{\partial f_A}{\partial {\phi}_A} + D_{12} \frac{\partial f_B}{\partial {\phi}_A} \right]\\
-\hat{\phi}_B \left[ D_{12} \frac{\partial f_B}{\partial {\phi}_B} + D_{21} \frac{\partial f_A}{\partial {\phi}_B} \right]
\label{cond_gen}
\end{multline}

Then a straightforward calculation results in the following compact conditions for the instability to occur: 
\begin{eqnarray}
\Gamma&>&0  \\
\Gamma^2 &>& 4 D_{11} D_{22} \left( 1 - \frac{D_{12}}{D_{11}} \phi_A - \frac{D_{21}}{D_{22}} \phi_B \right) \det(\mathbf{A}) 
\nonumber
\end{eqnarray}
together with $\tr(\mathbf{A})<0$ and $\det(\mathbf{A})>0$.

\begin{figure}[tb]
\begin{center}
\begin{tabular}{cc}
\phantom{x} & \phantom{x}\\
\includegraphics[scale=0.24]{fig4_FCD.eps}&
\includegraphics[scale=0.29]{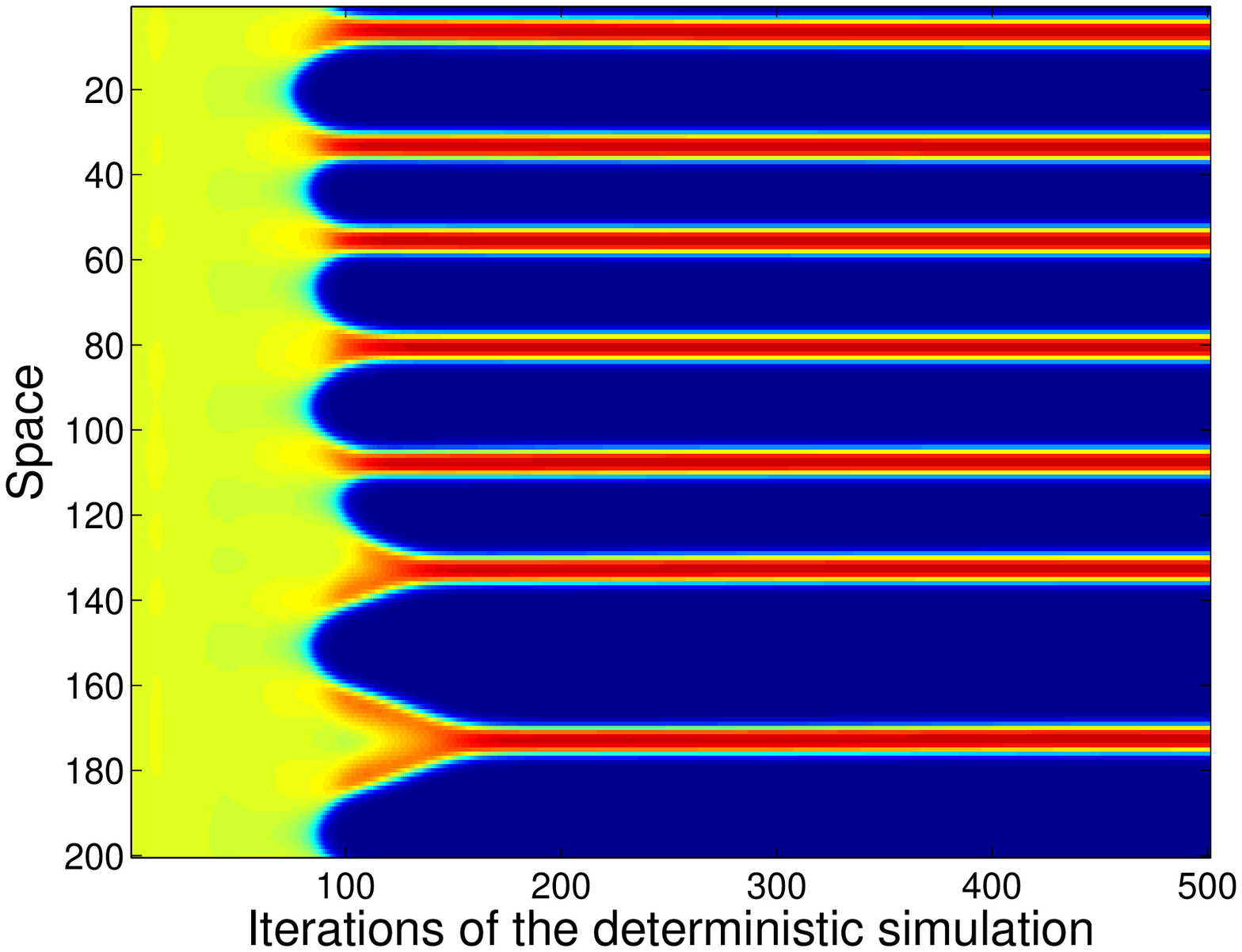}\\
(a) & (b)\\
\phantom{x} & \phantom{x}\\
\phantom{x} & \phantom{x}\\
\includegraphics[scale=0.24]{fig1_FCD.eps}& 
\includegraphics[scale=0.29]{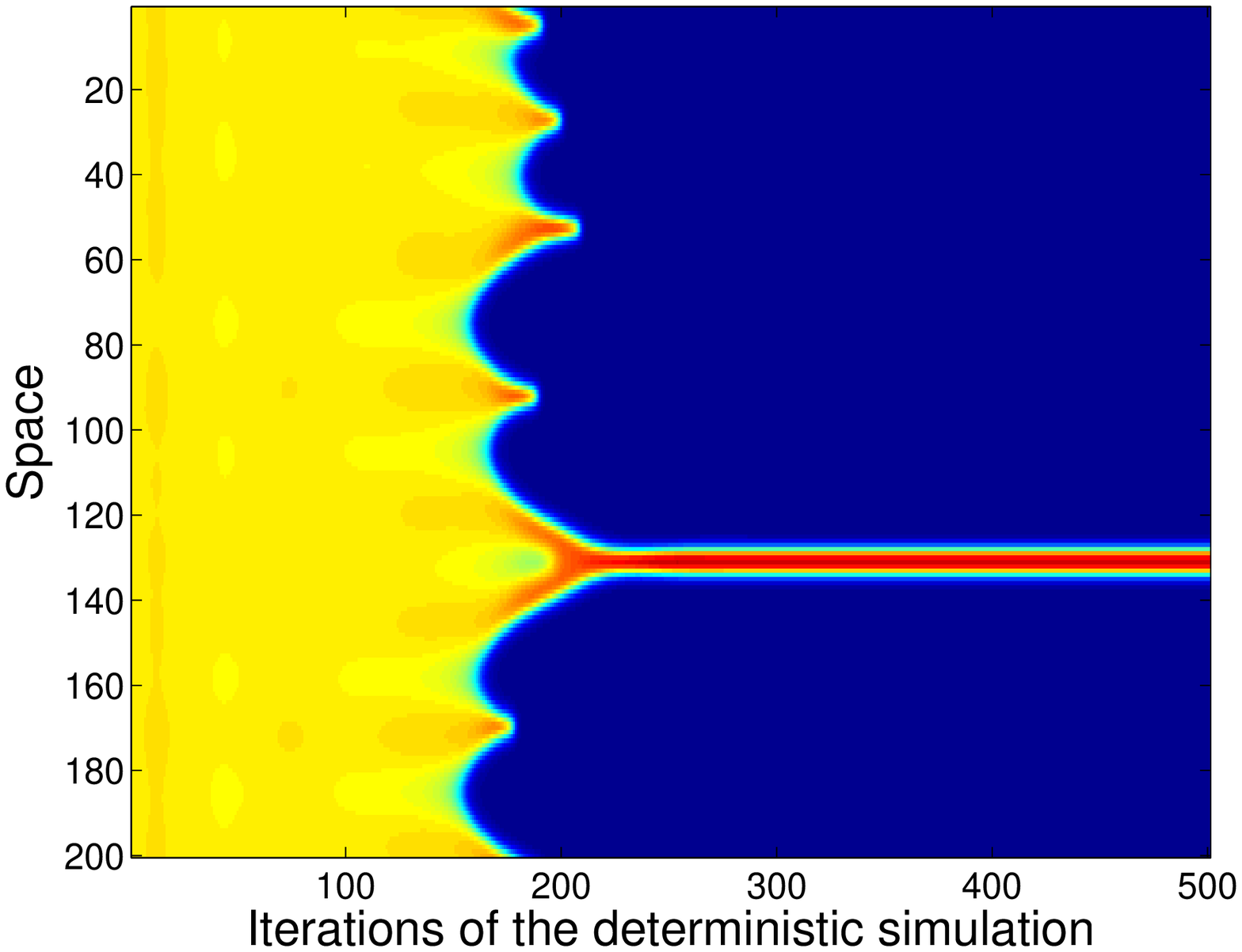}\\
\phantom{x} & \phantom{x}\\
(c) & (d)
\end{tabular}
\end{center}
\caption{Panels (a) and (c): the boundaries of the region of Turing instability are traced in the plane
($b,c$), for $D_{22}/D_{11} = 1$ (panel (a)) and $D_{22}/D_{11} = 0.7$ (panel (c)). 
The calculated domains refer to the Brusselator model with non Fickean diffusion, as 
explained in the main text.  The solid line, which encloses regions I and II, stands for $\Delta D=0$, while the dashed line delimits region I, where the condition $\Delta D=0.1$ applies. The other parameters are set as $a=5$, $d=3$. Panels (b) and (d):  the time evolution of the concentration $\phi_A$, as revealed by direct numerical simulations. In both cases, a small perturbation is superposed at $t=0$ to the (non trivial) stable homogeneous fixed point of the Brussellator, namely $\hat{\phi}_A = (a+\sqrt{a^2-4 a b (a+d) /c})/2/(a+d)$, $\hat{\phi}_B =b/c/\hat{\phi}_A$. Here, $D_{11}=1.0$, $D_{22}=0.7$, $b=21.71$, $c= 139$, $a=5$, $d=3$. The upper right figure, panel (b), refers to $\Delta D = 0$, the lower right, panel (d), to $\Delta D = 0.1$. In the simulations we have assumed a symmetric box $[-L,L]$, with $L=10$. The box is discretized in $200$, uniformly spaced, mesh points. The simulations are run by employing an explicit Euler scheme with time step equal to $0.0001$. The density in each cell of the mesh is displayed in the vertical axis, while the horizontal  axis refers to the number of iterations.}
\label{figure1}
\end{figure}

For demonstrative purposes we now specialize on a particular case study and trace out in the parameters' plane, the domain that corresponds to the Turing instability. Our choice is to work with the Brusselator model\footnote{The term $a (1-\phi_A-\phi_B)$ reflects the presence of the finite carrying capacity, as discussed in 
\cite{biancalani2010}. Similar conclusions hold however if the diluted limit is performed, {\it just} in the reaction terms, hence replacing $a (1-\phi_A-\phi_B)$ with $a$.} which implies setting $f_A=-(b+d) \phi_A + a(1-\phi_A-\phi_B) + c \phi_A^2 \phi_B$ and  $f_B= b \phi_A-c \phi_A^2 \phi_B$. Species $A$ plays now the role of the activator, while $B$ stands for the inhibitor. Results of the analysis are reported in left panels of Fig. \ref{figure1}, where the region of interest is singled out in the plane ($b,c$), for different choices of $\Delta D$. Turing patterns are predicted to occur for  $D_{22}/D_{11} \le 1$, at odd with what happens in the conventional scenario where standard Fick's diffusion is assumed to hold (see below). The right panels report the results of direct simulations and confirm the presence of macroscopically organized patterns in a region of the parameters space that is made classically inaccessible by the aforementioned, stringent condition $D_{22} > D_{11}$
The simulations refers to the choice $D_{22} / D_{11} =0.7$.
These observations are general and similar conclusions can be drawn assuming other reactions schemes of the inhibitor/activator type, different from the Brusselator model. 

\begin{figure}[tb]
\begin{center}
\includegraphics[scale=0.31]{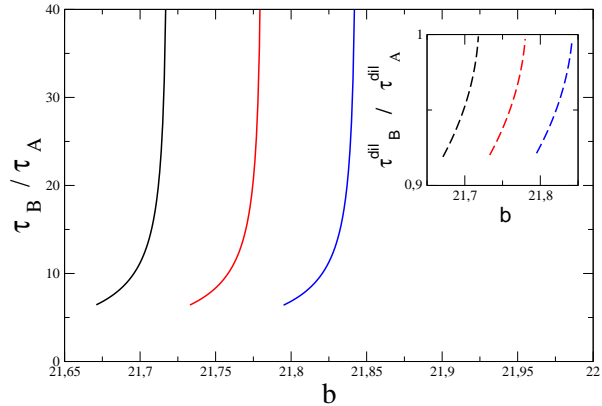} 
\end{center}
\caption{Main figure: the ratio $\tau_B/\tau_A$ is plotted for the Brusselator model, inside the Turing region, as a function of the chemical parameter $b$, for different choices of $c$. From left to right, $c=139, 139.4, 139.8$. The other parameters are set as in Fig. \ref{figure1}a, with $\Delta D=0$.   $\tau_B$ and $\tau_A$ follow respectively Eqs. (\ref{tauB}) and (\ref{tauA}) and quantify the time scales of  the reactive processes, within the framework of the generalized reaction diffusion scheme. As expected, the existence of a region of Turing order, as revealed in Fig \ref{figure1}a,  implies $\tau_B > \tau_A$. In the inset, the ratio of the time scales $\tau_B^{dil}/\tau_A^{dil}$ obtained in the diluted limit is reported and proven to be smaller than unit.  
}
\label{figure1b}
\end{figure}

It is now instructive to elaborate on a simple interpretation of the above result. Let us start by briefly revisiting the necessary conditions for the classical Turing instability to occur, namely:
\begin{eqnarray}
\tr(\mathbf{A}) = \partial f_A / \partial {\phi}_A + \partial f_B / \partial {\phi}_B &<& 0  \label{cond_class} \\ 
D_{11} \partial f_B / \partial {\phi}_B + D_{22} \partial f_A / \partial {\phi}_A &>& 0.  \nonumber
\end{eqnarray}
Both conditions can be simultaneously matched, only if the diagonal elements of the Jacobian matrix $\mathbf{A}$ have opposite signs. For the sake of clarity, let us assume\footnote{This is indeed the case for the Brusselator model. For $c$ sufficiently large, see also panels (a) and (c) of Fig. \ref{figure1}, 
we have in fact  $\frac{\partial f_A}{\partial {\phi}_A} \simeq 2 c  \hat{\phi}_A  \hat{\phi}_B  >0 $ and $\frac{\partial f_B}{\partial {\phi}_B} = -c  \hat{\phi}_A^2 <0 $.}
that: 
\begin{equation*}
\frac{\partial f_A}{\partial {\phi}_A}>0 \qquad \frac{\partial f_B}{\partial {\phi}_B}<0. 
\end{equation*}
Hence, species $A$ activates its own production, while species $B$ has a self-inhibitory feedback. Requiring $\tr(\mathbf{A})<0$ implies imposing 
$|\frac{\partial f_B}{\partial {\phi}_B}|> \frac{\partial f_A}{\partial {\phi}_A}$ which, by making use of the second of (\ref{cond_class}), readily translates 
into the necessary condition 
\begin{eqnarray}
\label{stand_cond_T}
\frac{D_{22}}{D_{11}} >  \frac{|{\partial f_B}/{\partial {\phi}_B|}}{{\partial f_A}/{\partial {\phi}_A}} > 1
\end{eqnarray}
As already mentioned, the inhibitor must diffuse faster than the activator (when the two species are evolved in separate containers) for the conventional Turing pattern to occur: the system has to accommodate for two competing processes, a short--range activation and long--range inhibition. Starting from this setting we can adapt the above reasoning to the generalized case study where cross diffusion terms are also present.  To this end, and to keep the notation light, we shall solely consider the limiting case with $\Delta D=0$. Similar conclusions hold when $\Delta D \ne 0$.  The second of relations (\ref{cond_class}) is now replaced by the condition $\Gamma >0$ (see Eq. (\ref{cond_gen})), which can be cast in the form:   
\begin{eqnarray}
\label{nec_cond_GT}
&& D_{11} \left[ \frac{\partial f_B}{\partial {\phi}_B} \left( 1- \hat{\phi}_B \right) -\hat{\phi}_A \frac{\partial f_B}{\partial {\phi}_A}\right]  \\
&+& D_{22} \left[ \frac{\partial f_A}{\partial {\phi}_A} \left( 1- \hat{\phi}_A \right) -\hat{\phi}_B \frac{\partial f_A}{\partial {\phi}_B}\right] >0 \nonumber
\end{eqnarray}
when $D_{11}=D_{12}$ and $D_{22}=D_{21}$. To proceed in the discussion we note that the  elements that enter the square brackets have dimension of the inverse of time. Assume $\frac{\partial f_B}{\partial {\phi}_B} \left( 1- \hat{\phi}_B \right) -\hat{\phi}_A \frac{\partial f_B}{\partial {\phi}_A}$ to be negative as it is reasonable to hypothesize if (i) the correction term that scales to the number densities $\hat{\phi}_A$ is sufficiently small, or conversely if (ii) we require $\partial f_B / \partial {\phi}_A > 0 $ (i.e. the first species stimulates with a positive feedback the other). Under these conditions, one can then introduce the characteristic time scale $\tau_B$  associated to the reaction dynamics of species $B$, defined as:
\begin{equation}
\tau_B = \left[ | \frac{\partial f_B}{\partial {\phi}_B} | \left( 1- \hat{\phi}_B \right) + \hat{\phi}_A \frac{\partial f_B}{\partial {\phi}_A} \right]^{-1}.
\label{tauB}
\end{equation}
Similarly, for species $A$, we have:
\begin{equation}
\tau_A = \left[  \frac{\partial f_A}{\partial {\phi}_A}  \left( 1- \hat{\phi}_A \right) - \hat{\phi}_B \frac{\partial f_A}{\partial {\phi}_B} \right]^{-1},
\label{tauA}
\end{equation}
assuming $\partial f_A / \partial {\phi}_A$ to control the sign in the above expression, or alternatively imposing $\partial f_A / \partial {\phi}_B <0$ (i.e. the second species acts with a negative feedback on the first one). The necessary condition (\ref{nec_cond_GT}) for the generalized Turing instability to occur takes the form:
\begin{equation*}
l_A^2 = \tau_A D_{11} < \tau_B D_{22} =l_B^2. 
\end{equation*}
where we have introduced two characteristic length scales, respectively $l_A$, $l_B$, associated to the reactive dynamics of species $A$ and $B$. In practice, also when $D_{22} < D_{11}$, spatially organized patterns can develop in the generalized reaction diffusion scheme provided the activator has a shorter life time, than the inhibitor. In formulae, $\tau_A = \tau_B D_{22} / D_{11} < \tau_B$ . In practical terms, the competition for the microscopic spatial resources modifies the time scales associated to the reactions processes and induces a self--consistent long--range effect that enlarges  the region of influence of the (isolated) inhibitors, also when the microscopic diffusion of the (isolated) activator is assumed to be faster. The crossed terms in the diffusion matrix determine a non trivial modification of the underlying characteristic times, which are now also sensitive to the off--diagonal elements of the Jacobian matrix.  In the diluted limit in fact, $\tau_A \rightarrow \tau_A^{dil} = (\partial f_A/ \partial {\phi}_A)^{-1}$ and 
$\tau_B \rightarrow \tau_B^{dil} =  | \partial f_B / \partial {\phi}_B|^{-1}$ and one is brought back to the standard, stringent condition (\ref{stand_cond_T}). In Fig. \ref{figure1b}  the ratio $\tau_B/\tau_A$ is displayed for the Brusselator model, inside the Turing region, as a function of the chemical parameter $b$. Different curves refer to distinct choices of $c$, while the other parameters are set to the values of Fig. \ref{figure1}a, with $\Delta D=0$. As expected, $\tau_B/\tau_A > 1$ a condition that eventually yields the generalized Turing patterns as described above. Conversely, and as pictured in the small inset, $\tau_B^{dil}/\tau_A^{dil}<1$. Hence, since $D_{22} < D_{11}$, Turing patterns cannot manifest via the classical pathway, which applies to diluted conditions. 

\begin{figure}[tb]
\begin{center}
\begin{tabular}{cc}
\includegraphics[scale=0.31]{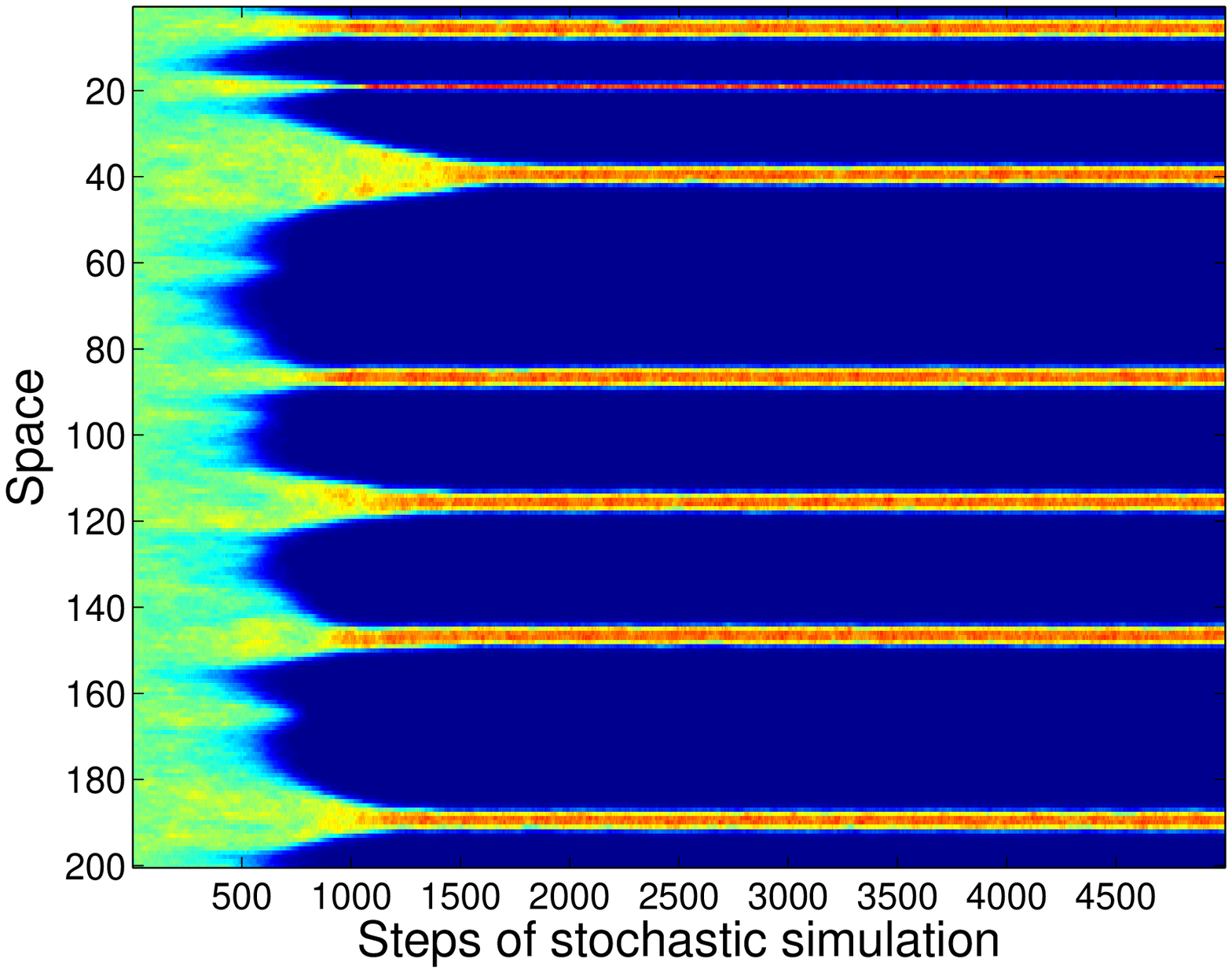}& 
\includegraphics[scale=0.31]{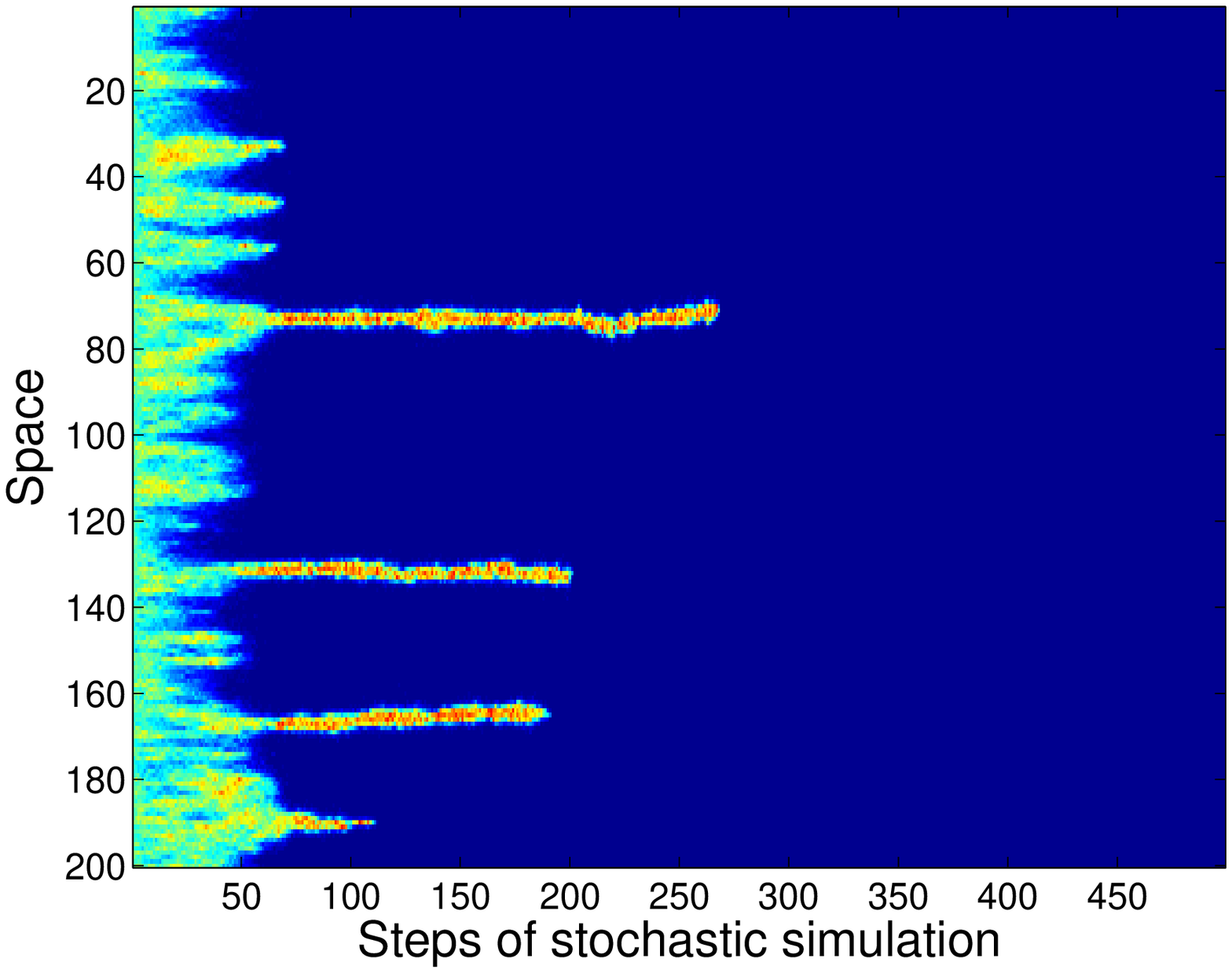}\\
(a) & (b)\\
\end{tabular}
\end{center}
\caption{
Time evolution of the discrete concentration $n_A/N$, as it results from a direct integration of the stochastic Brusselator model. The simulations follows the Gillespie algorithm \cite{gillespie}. Parameters refer to region II of Fig. \ref{figure1}b, namely  $D_{11}=1.0$, $D_{22}=0.7$, $\Delta D=0$, $a=5$, $d=3$, $b=21.71$,  $c=139$. In panel (a):  $N=3000$, while in panel (b)  $N=300$. Demographic fluctuations destroy the deterministic patterns which are hence interpreted as a metastable regime of the finite $N$ stochastic dynamics.}
\label{figure2}
\end{figure}

The remaining part of this section is devoted to discussing the robustness of the patterns depicted in Fig. \ref{figure1}  (panels (b) and (d)), and obtained upon integration of the governing system of partial differential equations.  It should be emphasized however that the model of multispecies diffusion here considered is stochastic in nature. It is therefore interesting to further elaborate on the contributions played by finite size effects, associated to the graininess of the system, and hence deliberately neglected under the idealized deterministic representation of the dynamics.  To this aim, one can carry out stochastic simulations, based on the Gillespie algorithm \cite{gillespie},  which produces realizations of the dynamics formally equivalent to those obtained from the governing master equation (\ref{master}).
We have here chosen to operate for the parameter setting of Fig. \ref{figure1}b and the results of our analysis are reported in Fig. \ref{figure2}. If the number of elements $N$ is sufficiently large ($N=3000$, in the left panel of Fig. \ref{figure2}) the patterns appear robust and resemble those recorded when operating in the framework of the deterministic picture. However, if the total number of microscopic individuals is reduced ($N=300$, in the right panel of Fig. \ref{figure2}) the patterns are less distinct and eventually fade away. Demographic fluctuations ultimately destroy the self-organized spatial patterns, relic of Turing instability, and the system evolves towards an asymptotically stable homogeneous solution. The lifetime of the metastable non homogeneous patterns increases with the system size and formally diverges in the thermodynamic limit $N \rightarrow \infty$. Waiting for a sufficiently large time, also the apparently stable density structures as displayed in Fig. \ref{figure2}a are expected to coalesce and smear out.  In other words, and intriguingly enough, the two limits for $N \rightarrow \infty$ and $t \rightarrow \infty$ do not commute. If the system size limit is taken before the infinite time limit, the dynamics is permanently frozen into a stationary non homogeneous configuration, the spatially ordered Turing patterns. Conversely,
the system is attracted towards a stable homogeneous equilibrium, due of the microscopic mixing that is seeded by the finite size fluctuations. Clearly the time of homogeneization can be extremely long, when compared to the finite time window of the experimental observation. In this respect, the metastable spatially extended patterns 
are possibly the solely regimes to be accessible to direct measures. This observation shares many similarities with the phenomenon of Quasi--Stationary States, so far associated to the long range nature of the two--body interaction \cite{Ruffo,Fanelli}. These findings, as well as the analysis of \cite{rogers}, can possibly shed new light onto the emergence of the Quasi--Stationary States, beyond the domain of applications for which they have been reported to occur. As a side remark, it is worth emphasising that similar conclusions hold when considering the diluted limit, i.e. when neglecting the role of a finite carrying capacity and the competition for the finite spatial resources that eventually yield the generalized cross diffusion terms here considered.

\section{Conclusions}
Summing up, Turing patterns can develop for virtually {\it any} ratio of the main diffusivities in a multispecies setting. This striking effect originates from the generalized diffusion theory that is here assumed to hold and that builds on the scheme discussed in \cite{fanellimckane}. Because of the competition for the available resources, a modified (deterministic) diffusive behaviour is recovered: cross diffusive terms appear which links multiple diffusing communities and which add to the standard Laplacian terms, relic of Fick's law. The fact that Turing like patterns are possible for e.g. equal diffusivities of the species involved\footnote{Notice that the authors of \cite{biancalani2010} failed to realize that accounting for cross diffusion terms of the type derived in \cite{fanellimckane} could result in an extension of the Turing mechanism to regions where $D_{22} \le D_{11}$. }, as follows a sound dynamical mechanism, constitutes an intriguing observation that hold promise to eventually reconcile theory and experimental evidences. The investigated setting applies in particular to multispecies systems that evolve in a crowded environment, as it happens for instance inside the cells where different families of proteins and other biomolecular actors are populating a densely packed medium. It is interesting to notice that the stochastic fluctuations, endogenous to the scrutinized system in its discrete version, eventually destroy the patterns, that are instead deemed to be stable according to the idealized deterministic viewpoint. The lifetime of the metastable patched patterns increases however with the size of the system, in striking analogy with what has been observed for the so called Quasi--Stationary States, out--of--equilibrium regimes observed in systems subject to long--range interactions. For large enough $N$, the homogeneization as seeded by fluctuations is progressively delayed and eventually prevented in the continuum limit $N \rightarrow \infty$.

\section{Acknowledgements}
We wish to thank Alan McKane and Tommaso Biancalani for useful discussion. The work is supported by Ente Cassa di Risparmio di Firenze and the program PRIN2009.

\end{document}